\documentclass{article}
\usepackage{ijcai26}

\usepackage{times}
\usepackage{soul}
\usepackage{url}
\usepackage[hidelinks]{hyperref}
\usepackage[utf8]{inputenc}
\usepackage[small]{caption}
\usepackage{graphicx}
\usepackage{amsmath}
\usepackage{amsthm}
\usepackage{booktabs}
\usepackage{algorithm}
\usepackage[switch]{lineno}
\usepackage{todonotes}

\usepackage{amsmath,amssymb,amsfonts}
\usepackage{algpseudocodex}
\usepackage{tikz}
\usetikzlibrary{matrix,positioning,calc}
\usepackage{subcaption}
\usepackage{xcolor} 
\usetikzlibrary{arrows.meta}   

\usepackage{url}

\newtheorem{theorem}{Theorem}

\title{Gradient-Based Join Ordering\thanks{To appear in the proceedings of 
IJCAI 2026, the 35th International Joint Conference on 
Artificial Intelligence.}}

\author{
Tim Schwabe
\And
Maribel Acosta\\
\affiliations
Technical University of Munich\\
\emails
\{tim.schwabe, maribel.acosta\}@tum.de
}

\begin{document}

\maketitle

\begin{abstract}
Join ordering is the NP-hard problem of selecting the most efficient order in which to evaluate joins (conjunctive, binary operators) in a database query. Because query execution performance critically depends on this choice, join ordering lies at the core of query optimization. Traditional approaches cast this problem as a discrete combinatorial search over binary trees guided by a cost model, but they have trade-offs between effectiveness and efficiency.
We show that when the cost model is differentiable, query plans can be continuously relaxed into a soft adjacency matrix that represents a superposition of plans. This continuous relaxation, combined with differentiable constraints that enforce plan validity, enables a gradient-based search for low-cost plans within this relaxed space. Using a Graph Neural Network as the cost model, we demonstrate that this gradient-based approach can find comparable and even lower-cost plans compared to traditional discrete search methods on two different graph datasets. Furthermore, we empirically show that the runtime of this approach scales better than discrete search algorithms. We believe this first step towards gradient-based join ordering can lead to more effective and efficient query optimizers in the future.
\end{abstract}

\section{Introduction}
\begin{figure}[t]
\centering
\includegraphics[width=0.45\textwidth]{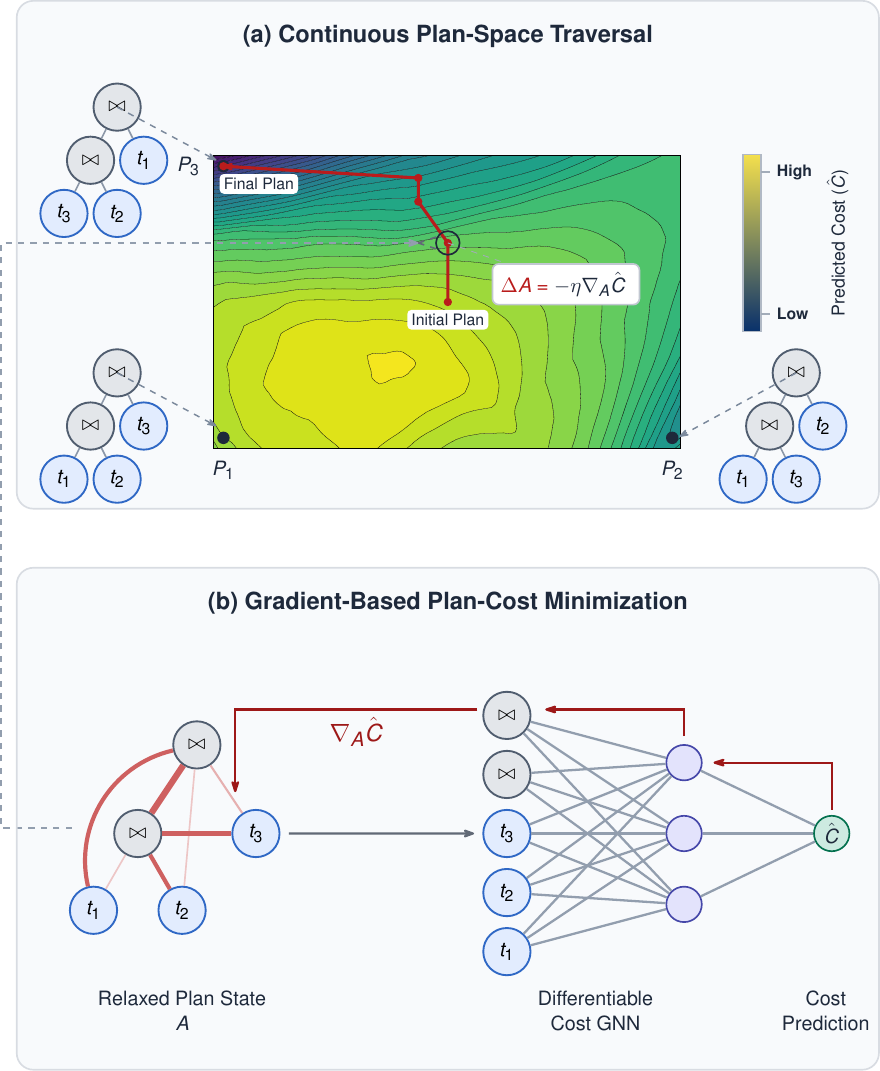} 
\caption{Join Ordering using a learned Cost Model. (a) Continuously relaxing the discrete Search Space of Plans allows traversing it using Gradient Descent. (b) The gradient of the cost with respect to the relaxed plan is backpropagated through the cost model.}
\label{fig:approach}
\end{figure}

Query optimization, particularly join ordering, is one of the most critical tasks in any database system. It involves determining the optimal evaluation order for the joins in a query, such that the resulting execution order has a low query runtime. Efficiently ordering join operations can significantly impact query execution time, often resulting in either instantaneous responses or impractically long delays.

At the core of join ordering is a \textit{cost model}, which estimates the computational expense associated with executing a query or a part of it. Accurately estimating these costs is a challenging statistical estimation problem, primarily due to the complex data distributions and correlations within real-world datasets. But, finding a good join ordering does not only involve estimating one specific order. It constitutes a discrete search problem over the space of plans, which are represented as binary trees, and which grows super-exponentially.
Classical approaches to navigating this space typically rely on discrete methods, such as dynamic programming, which systematically evaluates possible plans at an exponential cost, or heuristics, which choose locally optimal steps quickly but often find suboptimal solutions.

Recently, learned cost models based on neural networks have been proposed, demonstrating significant superiority in cost estimation compared to classical statistical methods~\cite{DBLP:conf/sigmod/ZhuWDZ24}. However, these learned cost models are either still paired with traditional discrete search methods or completely rely on Reinforcement Learning, which is sample-inefficient and has unstable convergence dynamics.

In this work, we present a fundamentally different approach: we can exploit the differentiability of learned cost models and reformulate join ordering as a continuous optimization problem by relaxing the discrete search space. As shown in Figure~\ref{fig:approach}, we represent query plans as continuous, i.e., edges are not binary but weighted between 0 and 1. This enables direct gradient-based search guided by a differentiable cost model. The main contributions of this work are:

\begin{itemize}
    \item We propose a novel gradient-based method for join ordering using learned cost models based on Graph Neural Networks, operating on a continuous relaxation of the discrete search space.
    \item We empirically demonstrate that this method generates solutions that match and sometimes surpass those found by discrete search methods.
    \item We show that the runtime of this approach is lower than typical randomized discrete baselines, and scales better than many of the classic discrete approaches.
\end{itemize}

\section{Preliminaries}
\label{sec:preliminaries}

In this work, we assume a graph data model, where queries typically involve a large number of joins due to the fine-grained, triple-based representation of relationships.
We assume basic familiarity with sets, graphs, and matrix notation. 
\subsection{Triple‑Based Graph Queries}
\label{sec:triple-queries}

We assume a graph data model in which the data is stored as
\emph{triples}, i.e. labelled, directed edges
\[
t = (s,p,o)\in\mathcal{E},
\]
where each element belongs to a set
$s,p,o\in\mathcal{I}$.
A finite set of triples $\mathcal{E}$ forms an
edge‑labelled directed graph.

\vspace{2pt}\noindent
\textbf{Triple patterns and conjunctive queries.}
Such graphs are queried with \emph{triple patterns}, i.e., triples in which any position may be a variable from an infinite set
$\mathcal{V}$ (s.t. $\mathcal{V} \cap \mathcal{I}=\emptyset$) of \emph{variables}:
\[
tp=(s,p,o)\in(\mathcal{I}\cup\mathcal{V})^3.
\]
A triple pattern is the most basic query.  Its answer is the set of
solution mappings
\[
\Omega_{tp}
=\bigl\{\mu:vars(tp)\!\to\!\mathcal{I}\,\bigm|\,
      \mu(tp)\in\mathcal{E}\bigr\},
\]
where $vars(tp)$ are the variables appearing in $tp$.
The number of solution mappings is called the \emph{cardinality} $|\Omega_{tp}|$ of the triple pattern.
More general \emph{conjunctive queries} combine several triple patterns
\(
Q=\{tp_1,\dots,tp_n\}\subseteq(\mathcal{I}\cup\mathcal{V})^3
\).

\vspace{2pt}\noindent
\textbf{Join Operators.}
The answer to a conjunctive query is obtained by iteratively \emph{joining} the solution mappings. Let
$\operatorname{dom}(\mu)$ denote the variables on which
$\mu$ is defined.  
Mappings $\mu_1,\mu_2$ are \emph{compatible}, $\mu_1 \sim \mu_2$ if they assign the same
value to every shared variable, i.e.
$\forall v\in\operatorname{dom}(\mu_1)\cap\operatorname{dom}(\mu_2):
\mu_1(v)=\mu_2(v).$
Their union $\mu_1\cup\mu_2$ is the mapping with domain
$\operatorname{dom}(\mu_1)\cup\operatorname{dom}(\mu_2)$.  
The join of two solution sets is then
\[
\Omega_1 \Join \Omega_2
=\bigl\{
      \mu_1\cup\mu_2 \ \big|\
      \mu_1\in\Omega_1,\;
      \mu_2\in\Omega_2,\;
      \mu_1 \sim \mu_2
  \bigr\},
\]
i.e., a set of solution mappings with cardinality $|\Omega_1\Join\Omega_2|$.





\subsection{Query Plans}

The final set of solution mappings is invariant to the order in which the triple patterns are joined. However, since the number of intermediate results can differ vastly, the order still matters. 
A \emph{query plan} $p$ for $Q$ defines the order in which these joins occur. Formally, it is a binary tree whose leaves correspond exactly to the triple patterns in \(Q\), and internal nodes represent join operators combining intermediate results from their children.

\begin{itemize}
    \item A plan is \emph{bushy} if no structural restrictions exist. Joins may freely combine the results of arbitrary subplans.
    \item A plan is \emph{left-linear} if each join operator has at most one non-leaf child. Such plans represent sequential join orders.
\end{itemize}

\begin{figure}[t]
\centering
\includegraphics[width=0.45\textwidth]{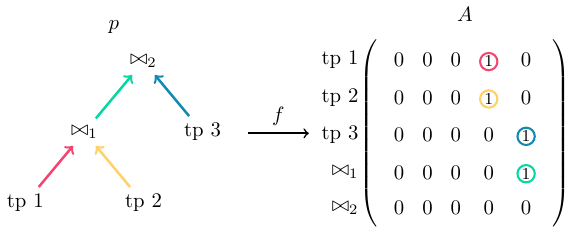}
\caption{A join plan $p$ is represented as a $\left(2n-1\right) \times \left(2n-1\right)$ dimensional matrix $f(p)=A$. The first $n$ rows denote the outgoing edges of triple patterns, while the last $n-1$ rows are outgoing edges of join nodes. The root node always corresponds to the last row.}
\label{fig:plan_representation}
\end{figure}

Since the space of bushy plans is vast ($C_{n-1}n!$, where $C_{n-1}$ is the Catalan Number), commercial optimizers often restrict the search space to left-linear plans (with size $n!$). While our proposed method is general and can search the space of bushy plans, we first focus on left-linear plans in this work.
For a query with $n$ triple patterns, we represent a plan $p$ as an adjacency matrix \(A\in\{0,1\}^{\left(2n-1\right) \times \left(2n-1\right)}\), where \(A_{ij}=1\) means node \(i\) is joined into node \(j\). The first $n$ indices in $A$ correspond to triple patterns, and the last $n-1$ indices correspond to join nodes (see example in Figure~\ref{fig:plan_representation}).

\subsection{Execution Cost and Cost Model}

The goal of join ordering is to minimize the execution \emph{cost} of a plan, which measures resources used, e.g., runtime, memory, disk access. Formally, the cost is defined as a function
$C:\mathcal{P}\rightarrow\mathbb{R}_{\geq0}$,
mapping a plan $p \in \mathcal{P}$ to its cost. 

During optimization, a \emph{cost model} is employed to approximate the true cost:
\[
\hat{C}_\theta:\mathcal{P}\rightarrow\mathbb{R}_{\geq0},
\]
with parameters \(\theta\). Simpler approaches rely on nonparametric statistical synopses, while recent research often models \(\hat{C}_\theta\) using neural networks.

\subsection{Cost Definition (\(C_{\text{\mdseries\upshape out}}\))}

There are many options to define the cost of a query plan and its internal join operations. A popular definition of cost that is widely used and adopted in this paper is the so-called $C_\text{out}$ cost function~\cite{DBLP:conf/icdt/CluetM95}. It recursively defines the cost of a join $v$ as its output cardinality plus the costs of the left ($l$) and right ($r$) sides of the join:
\[
C_{\text{out}}(v) = |\Omega(v)| + C_{\text{out}}(l) + C_{\text{out}}(r),
\]
The cost of a leaf node (a triple pattern) is defined to be 0:
\[
C_{\text{out}}(tp_i)=0,\quad \text{for all leaves } tp_i.
\]

\subsection{Discrete Search Algorithms for Join Ordering}\label{sec:baseline_methods}

In order to find a good join ordering, several search strategies are used. 
\textbf{Exhaustive Search (ES)} evaluates all possible plans and is guaranteed to find the optimum of the given cost model. However, its factorial complexity (\(\mathcal{O}(n!)\) for left-linear plans) makes it impractical beyond small queries.

\textbf{Dynamic Programming (DP)}~\cite{DBLP:conf/sigmod/SelingerACLP79} reduces this to \(\mathcal{O}(3^n)\) by applying Bellman’s principle of optimality: the best plan for a set of joins can be built from optimal subplans. However, this only holds when the cost model satisfies this property and has no guarantees for arbitrary, non-monotonic cost models.

To circumvent the intractable time complexity of DP and ES, most optimizers employ heuristic search methods.
\textbf{Greedy Search}~\cite{DBLP:conf/dexa/Fegaras98} builds plans incrementally by choosing locally optimal moves. At every iteration, the triple pattern (or
intermediate sub‑plan) whose addition yields the lowest
estimated cost according to the cost model~$\hat C_\theta$ is appended to the join plan. Its downside is that it can find suboptimal local optima. However, it only requires \(O(n^2)\) evaluations of the cost model, making it computationally efficient. Several randomized algorithms have been explored and are routinely used in database systems. \textbf{Iterative Improvement (II)}~\cite{DBLP:conf/sigmod/Swami89} starts from a random initial plan and explores a local neighborhood (e.g., pairwise swaps of join positions). At each iteration, all swap neighbors (or a random sample) are evaluated, and if any neighbor improves the cost, the search moves to the best one. This continues until no improving neighbor exists (a local optimum).

\textbf{Genetic Query Optimization (GEQO)} applies evolutionary algorithms to join ordering~\cite{DBLP:journals/vldb/SteinbrunnMK97}. It is implemented in PostgreSQL as the default optimization strategy for queries with more than 12 joins~\cite{postgres_geqo_intro}. A population of candidate plans (represented as permutations) evolves over generations through selection (ranked by predicted cost), crossover (edge-recombination crossover~\cite{whitley1989scheduling}), and mutation operators (e.g., swapping two join positions, as in II).


\section{Related Work}

\textbf{Learned Cost Models}\\[2pt]
Traditional query optimizers rely on formula-based cost estimators that combine histograms, independence assumptions, and hand-crafted heuristics or sampling-based approaches, and are inaccurate for larger joins. 
Recent research, hence, turned to learned models to estimate the cardinality or cost of a query. These models try to map from a suitable representation of a query or plan to its cardinality or cost.

For relational data, \textit{MSCN}~\cite{MSCN} models tables, joins, and predicates with separate subnetworks that are aggregated to produce a cardinality estimate. To capture the inherent tree-based structure of query plans, several works use architectures such as Tree-LSTM~\cite{DBLP:journals/pvldb/SunL19}, or Transformers~\cite{QueryFormer} with tree-based attention.

For graph databases, several approaches have used Graph Neural Networks (GNNs)~\cite{gnce,LSS} or Recurrent Neural Networks~\cite{aytimur2025space} and pretrained embeddings to accurately represent queries and estimate their cardinality. Similarly to approaches on relational databases, approaches for cost estimation explicitly represent the tree structure of the query plan~\cite{casals2023sparql,mohanaraj2025planrgcn,qin2025neuro}. Building on this prior work, we use a GNN to estimate the cost of a plan.

\vspace{4pt}\noindent
\textbf{Learned Query Optimization}\\[2pt]
The learned cardinality and cost estimators above can be used to search for optimal plans using discrete search methods. To combine the time efficiency of greedy approaches with the ability to find lower-cost plans of more exhaustive approaches, recent work has turned to reinforcement learning. The search for an optimal plan is cast as a Markov decision problem where the optimal next action is joining two subplans (where a subplan can be an unjoined table or triple pattern) such that the expected cost is minimized. Various approaches in relational and graph databases use common algorithms such as policy gradients~\cite{ReJoin} or Q-Learning~\cite{JOGGER,RTOS}. Similar to learned cost models, those approaches use Tree-based representations~\cite{RTOS,NEO} and Graph Neural Networks~\cite{JOGGER,LOGGER} to represent the plans. Several surveys summarize the recent work in relational~\cite{DBLP:journals/pvldb/YanUL23,DBLP:conf/sigmod/ZhuWDZ24} and graph databases~\cite{Acosta2025}.

Since RL approaches typically suffer from data-inefficient training, our approach focuses on supervised learning, followed by a novel way to search the space of plans.

\vspace{4pt}\noindent
\textbf{Gradient Based Optimization of Discrete Structures}\\
Using Gradient Descent to optimize structures that are discrete in nature has been explored in various previous works and domains. \textit{DARTS}~\cite{DARTS} and \textit{SNAS}~\cite{SNAS} reformulate Neural Architecture Search to a smooth loss and use gradient descent to directly optimize the network architecture. The approach of \textit{DART} is transferred to differentiable program synthesis by \citeauthor{diff_program_synthesis}. \textit{NRI}~\cite{NRI} simultaneously learns the underlying discrete interaction graph and the dynamics model of a dynamical system using GNNs inside an autoencoder.
Structure learning, the task of finding the graph that best explains observed data, has also been approached using gradient‑based optimization~\cite{NoTears,golem,DBLP:journals/csur/VowelsCB23}.
In the space of control theory using learned world models, \citeauthor{DBLP:journals/corr/abs-2312-17227} perform gradient-based planning using a recurrent world model. In comparison to our work, they use gradient-based search to find a good order of actions to take, while we aim to find a good order of join operations.
Another line of work has applied continuous relaxation to the discrete process of sorting~\cite{NeuralSort,LecorveVBR22,SoftSort}. Neural Sort~\cite{NeuralSort} proposes a differentiable relaxation of \texttt{argsort} with a soft permutation matrix. 

Despite these diverse approaches to gradient-based optimization in discrete search spaces, their application to query optimization remains unexplored, a gap that we aim to address in this work.
\section{Gradient-Based Join Ordering}
\label{sec:approach}

We now present GBJO, our approach that reformulates join ordering using gradient descent.
We first formalise the optimization objective, then show how a
continuous relaxation of the discrete plan space enables end-to-end
gradient descent. 

\subsection{Optimization Objective}
\label{sec:opt_objective}

Given the set $\mathcal P(Q)$ of possible plans for a query
$Q$, i.e. $\mathcal{P}(Q) \subseteq \mathcal{P}$, and let
\(f:\mathcal P(Q)\!\to\!\{0,1\}^{N\times N}\)
map a plan \(p\) to its binary adjacency matrix
\(A = f(p)\), where \(N = 2n - 1\) is the total number of plan
nodes.
Given a feature matrix \(X\in\mathbb R^{N\times F}\) with $F$-dimensional features for the $N$ leaves, and a differentiable cost model \(\hat C_\theta\), join‑ordering is the following discrete optimization problem:

\begin{equation}
  \min_{p \in \mathcal P(Q)}
      \;\; \hat C_\theta\bigl(X,A).
  \label{eq:discrete_obj}
\end{equation}

\subsection{Continuous Relaxation of Plans}
\label{sec:cont_relax}

To transform the discrete objective in Equation~\eqref{eq:discrete_obj} into a continuous optimization problem, we relax the binary constraints on the adjacency matrix $A\in\{0,1\}^{N\times N}$. We introduce a real-valued \emph{logit matrix} $L\in\mathbb R^{N\times N}$ and derive edge probabilities by applying a  softmax:
\begin{equation}
    A^{\mathrm{soft}}_{ij} = \frac{\exp(L_{ij}/\tau)}{\sum_{k} \exp(L_{ik}/\tau)},
\label{eq:gumbel_sigmoid_short}
\end{equation}

$\tau \in \mathbb{R}^+$ is a temperature parameter and controls the sharpness of the distribution. $A^{\mathrm{soft}}$ now represents a continuous superposition of discrete plans. 

Since valid join plans are located at the binary vertices of the hypercube $[0,1]^{N\times N}$, the optimization must eventually drive each $A^{\mathrm{soft}}_{ij}$ towards either~$0$ or~$1$. This is achieved by annealing the temperature parameter $\tau$ from $\tau_0$ to a minimum value $\tau_{\min}$. As $\tau \to 0$, the entries of $A^{\mathrm{soft}}_{ij}$ approach binary values. To balance free exploration in early iterations with convergence towards discrete solutions in later stages, we linearly anneal the temperature over $I$ optimization steps like $\tau_t=\max \bigl(\tau_{\min},\,\tau_0-\tfrac{t}{I}(\tau_0-\tau_{\min})\bigr)$.

\vspace{4pt}\noindent
\textbf{Relaxed Optimization Problem.}
Relaxing the plan adjacency to continuous values now gives rise to a smooth
objective
\begin{equation}
  \min_{L\in\mathbb R^{N\times N}}
  \hat C_\theta\bigl(X, A^{\mathrm{soft}}\bigr),
  \label{eq:relaxed_obj_final}
\end{equation}
which can be approached using gradient-based methods.

\subsection{Structural Constraints}
\label{sec:constraints}

Optimizing Equation \eqref{eq:relaxed_obj_final} alone empirically
converges to degenerate minima, e.g., cyclic or disconnected
graphs, that do not correspond to valid query plans.
We therefore move to a constrained optimization problem by
adding differentiable penalties that quantify the invalidity of
a binary $A$ and vanish \emph{iff} it is a valid,
binary join tree.

\vspace{4pt}\noindent
\textbf{(i) Degree constraints.}
We want to enforce that triple-pattern nodes have exactly one outgoing edge,
while join nodes have exactly two incoming edges and one outgoing edge.
Furthermore, the (single) root join node has no outgoing edges.
Without loss of generality, let $r=2n-1$ be the index of the root node.
With in-degree $d^{\mathrm{in}}_v=\sum_i A^{\mathrm{soft}}_{iv}$
and out-degree $d^{\mathrm{out}}_v=\sum_j A^{\mathrm{soft}}_{vj}$,
we introduce quadratic penalties that are \emph{zero iff} the
degree conditions above are satisfied:
\begin{align*}
  P_{\textsc{to}} &= \sum_{v\leq n}\bigl(d^{\mathrm{out}}_v-1\bigr)^{2}, \ \ \   P_{\textsc{ji}} = \sum_{v>n}\bigl(d^{\mathrm{in}}_v-2\bigr)^{2},\\[2pt]
  P_{\textsc{jo}} &= (d^{\mathrm{out}}_r)^{2} + \sum_{n< v<r}\bigl(d^{\mathrm{out}}_v-1\bigr)^{2}.     
\end{align*}
Note that we do not require a penalty for the in-degree of the triple pattern, as in a valid join-tree, they never have incoming connections. Instead, all connections from any node to a triple pattern are set to negative infinity in $L$ and hence are 0 after applying the softmax. Similarly, since a valid root node can never have outgoing connections, we mask out all outgoing edges from it.

\vspace{4pt}\noindent
\textbf{(ii) Left‑linearity.}
Our approach supports any bushy tree. However, it can be enforced to produce left-linear plans. Let \(J=\{n+1,\ldots,2n-1\}\) denote the ordered set of join nodes. We require, without loss of generality  
(i) the first join node $j_0$ (index $n+1$) to have two triple‑pattern children and no
join child, and  
(ii) every subsequent join node \(v\in J\!\setminus\!\{j_0\}\) to have exactly
one triple‑pattern child and one join child.
With the sum of incoming connections from triple patterns $c^{\text{tp}}_v$ and from join nodes $c^{\text{jn}}_v$ to $v$ as
\[
  c^{\text{tp}}_v \;=\; \sum_{u\leq n}     A^{\mathrm{soft}}_{uv},
  \qquad
  c^{\text{jn}}_v \;=\; \sum_{u > n}  A^{\mathrm{soft}}_{uv},
\]
the left‑linear penalty becomes
\begin{equation*}
  P_{\textsc{ll}}
  \;=\;
  \bigl(c^{\text{tp}}_{j_0}-2\bigr)^{2}
  \;+\;
  \bigl(c^{\text{jn}}_{j_0}\bigr)^{2}
  \;+\;
  \sum_{v\in J\setminus\{j_0\}}
  \bigl[(c^{\text{tp}}_v-1)^{2} + (c^{\text{jn}}_v-1)^{2}\bigr].
  \label{eq:leftlinear}
\end{equation*}
The penalty is zero iff the plan is left‑linear.

\vspace{4pt}\noindent
\textbf{(iii) Acyclicity.}
Any valid join tree does not contain cycles. \citeauthor{NoTears} proposed a differentiable objective to measure the DAG-ness of a graph and hence we use the following penalty to suppress cycles:
\begin{equation}
  P_{\textsc{acyc}}\;=\;\operatorname{tr}\bigl(e^{A^{\mathrm{soft}}}\bigr)-N,
  \label{eq:acyclic}
\end{equation}
which is zero exactly if no cycles are present.

\noindent
Together, these penalties characterize a valid plan:
\begin{theorem}[Validity of the structural penalties\footnote{The proof is provided in the appendix.}]
\label{thm:penalties}
Let $n \geq 2$ and $A \in \{0,1\}^{N \times N}$ with $N=2n-1$, subject
to the masking introduced above: $A_{iv}=0$ for all $v \leq n$ and
$A_{rj}=0$ for all $j$, with $r = N$.
Then $P_{\textsc{to}}+P_{\textsc{ji}}+P_{\textsc{jo}}+P_{\textsc{ll}}
+P_{\textsc{acyc}} = 0$ if and only if $A$ is the adjacency matrix of
a valid left-linear plan.
\end{theorem}

\subsection{Constrained Objective and Optimization}
\label{sec:full_loss}

Let
$
  P_{\textsc{struct}}
  =
 \lambda_{\textsc{to}}P_{\textsc{to}}
 +\lambda_{\textsc{ji}}P_{\textsc{ji}}
  +\lambda_{\textsc{ll}}P_{\textsc{ll}}
 +\lambda_{\textsc{jo}}P_{\textsc{jo}}
 +\lambda_{\textsc{acyc}}P_{\textsc{acyc}}
$
be the weighted sum of all structural penalties.  The final time-dependent objective to
be minimized is now given as
\begin{equation}
  \mathcal L_t
  \;=\;
  \hat C_\theta\bigl(X, A^{\mathrm{soft}}\bigr)
  \;+\;
  \lambda(t)\,P_{\textsc{struct}},
  \label{eq:total_loss}
\end{equation}
where $\lambda(t)=\lambda_{\max}(t/I)^q$ nonlinearly increases
the influence of the constraints over the $I$ optimization steps, up to a parameter $\lambda_{\max}$. Here, the exponent $q$ is a hyperparameter. Similar to
$\tau$, this allows for largely unconstrained exploration early on and forces valid plans
at the end of optimization, and was empirically found to be critical for convergence.

The differentiability of the cost model, penalties and relaxed
adjacency matrix makes $\mathcal L_t$
amenable to standard gradient descent updates:
\begin{equation}
      L \;\leftarrow\; L - \alpha \,\nabla_L \mathcal L_t.
\end{equation}

We use GD with Nesterov momentum together with a One Cycle LR schedule for optimization. This enables fast convergence with a minimal number of gradient descent steps.


\paragraph{Projection to a Discrete Plan.}
To convert the best found soft adjacency \(A^{\mathrm{soft}}\) to a valid
left‑linear plan $p^\star$, we perform a simple greedy search:
with $T_{free}=\{1,\ldots,n\}, J_{free}=\{n+1,\ldots,2n-1\}$, starting from the root join \(c=2n-1\) we iteratively
select
\[
  (j^\star,k^\star)=\arg\max_{j\in J_{\text{free}},\,k\in T_{\text{free}}}
                    \bigl(A^{\mathrm{soft}}_{jc}+A^{\mathrm{soft}}_{kc}\bigr),
\]
and attach the edges \(j^\star\to c\) and \(k^\star\to c\) to $p^\star$, then set
\(c\leftarrow j^\star\) and $T_{free}=T_{free}\!\setminus\!k^\star, J_{free}=J_{free}\!\setminus\!j^\star$; the final join receives the two remaining
triples. This projection is guaranteed to yield a valid plan. For the final returned solution, we dynamically retain the best discretized plan encountered so far during optimization. The full pseudocode is provided in the appendix.

\section{Experimental Results}
\label{sec:experimental_results}
\subsection{Datasets and Queries}
We evaluate GBJO on two popular knowledge graphs: LUBM~\cite{LUBM} and a subset of 
 Wikidata (Wikidata 5M~\cite{Wikidata}). LUBM is a synthetic dataset used for Database Benchmarking, while Wikidata is one of the largest publicly available knowledge graphs. LUBM is a rather structured dataset, while Wikidata is complex and heterogeneous. For queries, we use the star- and path-shaped queries provided by \citeauthor{gnce} and generate additional larger queries using their query generator. Star queries have a single central node that connects to various outer nodes, while a path query is a single chain of nodes.
We selected these datasets and queries as they represent both smaller, simpler graphs as well as larger, more complex ones. The query shapes reflect those commonly studied in prior research and cover a wide range of cardinalities and costs, thus providing a representative range of typical query workloads.
\subsection{Plan Generation}
To train and evaluate the cost model, we need to obtain query plans for the used queries. A trivial way is to generate random join orders. However, this mostly generates plans with mediocre performance and does not provide plans with either very good or very poor performance. Hence, for each query, we generate three random, three good, and three bad plans. To generate good and bad plans, we use a bottom-up beam search with a beam width of three and direct execution on the database. For the good plans, we retain the three cheapest subplans (as per the true $C_{\text{out}}$ cost) at each step of the search, while for the bad plans, we retain the three worst plans.
\footnote{All used and generated datasets, as well as the code with instructions on how to reproduce the results, are available on GitHub: \url{https://github.com/TimEricSchwabe/GBJOv2}}

To represent the feature matrix $X$ of a plan, we again follow the scheme of~\citeauthor {gnce} and use Knowledge Graph Embeddings.

\subsection{Cost Model Training}\label{sec:cost_model_training}
Our approach is agnostic of model architecture and only expects the model to be compatible with a soft adjacency matrix. We choose a simple GNN with 6 message-passing layers and two fully connected layers for graph-level readout, transforming the initial plan graph into a single cost estimate (architecture details are given in the appendix). As the message-passing function, we use \textit{GIN Conv}~\cite{DBLP:conf/iclr/XuHLJ19} and extend it with edge weights in order to be able to represent the soft adjacencies. The GNN is trained on valid plans with hard adjacencies only to predict the $C_{out}$ cost.

The number of plans used for training and validation along with the median Q-error ($\max\bigl(\tfrac{\hat{C}}{C},\tfrac{C}{\hat{C}}\bigr)$) between true ($C$) and predicted ($\hat{C}$) cost is shown in the table below. 
\begin{center}
\small            
\setlength{\tabcolsep}{4pt}   
\begin{tabular}{lcccc}
\textbf{} & \textbf{LUBM–Star} & \textbf{LUBM–Path} & \textbf{WD–Star} & \textbf{WD–Path}\\
\# Train        &   61224   &  37304    &   72000    &    35229  \\
\# Val             &   15311   &   9326   &   18000    &    8807  \\
Q-error $\downarrow$    & 1.13 & 1.07 & 1.40  & 2.34 \\  
\end{tabular}
\end{center}

For both LUBM and Wikidata, and both query types, the models are accurately estimating the cost.

\subsection{Join Order Optimization}
We use the cost models trained in the previous step to compare GBJO against the baseline approaches presented in Section~\ref {sec:baseline_methods}, as well as two approaches operating on the continuous plan space: NeuralSort~\cite{NeuralSort} and CMA-ES~\cite{DBLP:journals/ec/HansenO01}. CMA-ES is a gradient-free evolutionary approach that optimizes continuous functions by adapting a multivariate normal search distribution over (soft) plans based on the covariance matrix of the more successful samples.
All compared search methods use the same model. Most of the hyperparameters are rather robust to changes in their values (as determined using hyperparameter search). The most important hyperparameter was to gradually increase the penalties as opposed to them being applied completely from the beginning (Equation~\ref{eq:total_loss}). For the approaches that require a set number of iterations, we set: $I_{\text{GBJO}}=10; \ I_{\text{II}} = 500; \ I_{\text{GEQO}} = 500; \ I_{\text{CMA-ES}} = 1500; \ I_{\text{NeuralSort}} = 10$.
\subsubsection{Visualizing the Cost Landscape}
\begin{figure}[t!]
  \centering
  \begin{subfigure}[b]{0.48\linewidth}
    \centering
    \includegraphics[width=\linewidth]{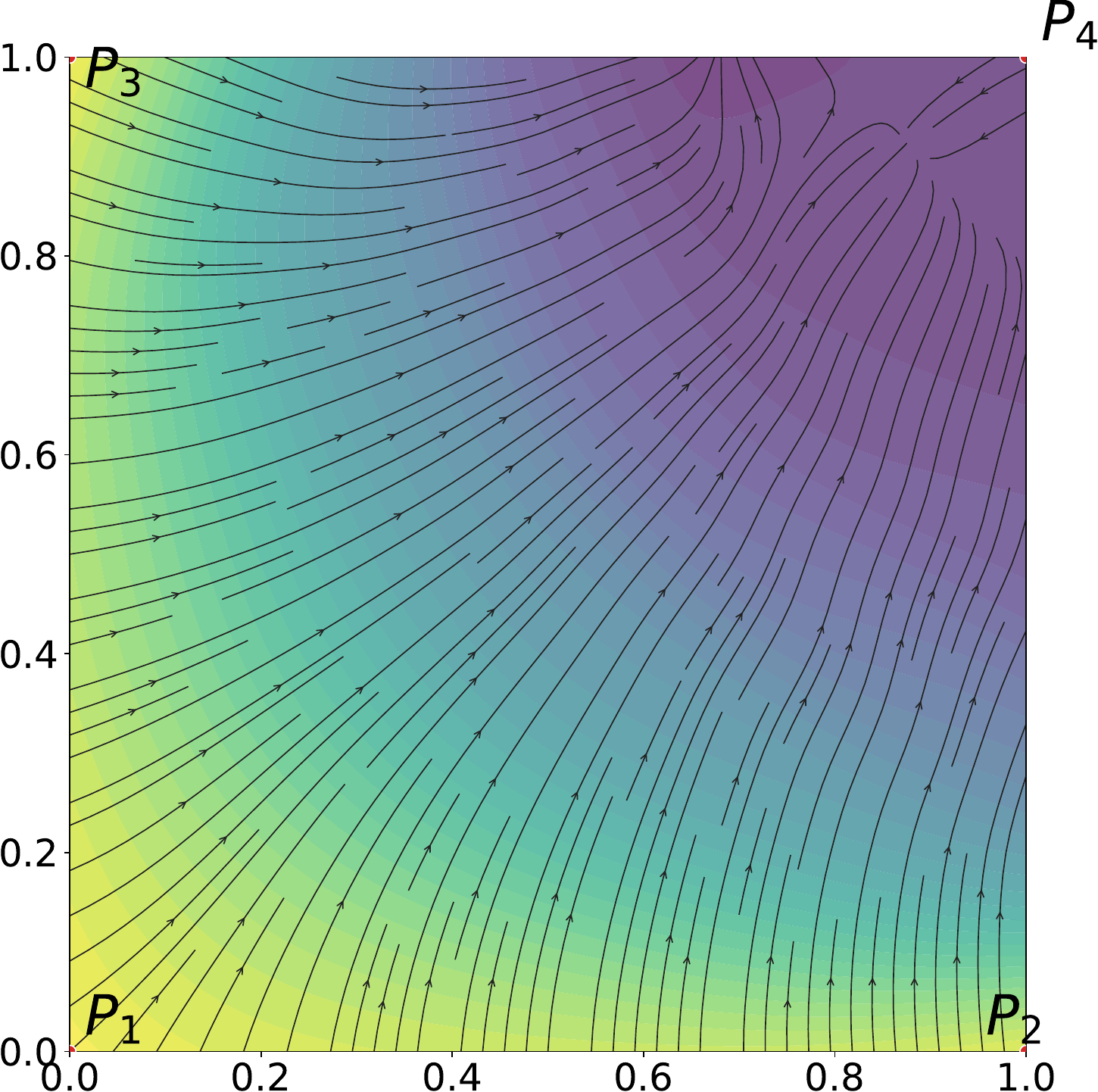}
    \caption{8 tp}   
    \label{fig:cost_landscape:a}
  \end{subfigure}
  \hfill
  \begin{subfigure}[b]{0.48\linewidth}
    \centering
    \includegraphics[width=\linewidth]{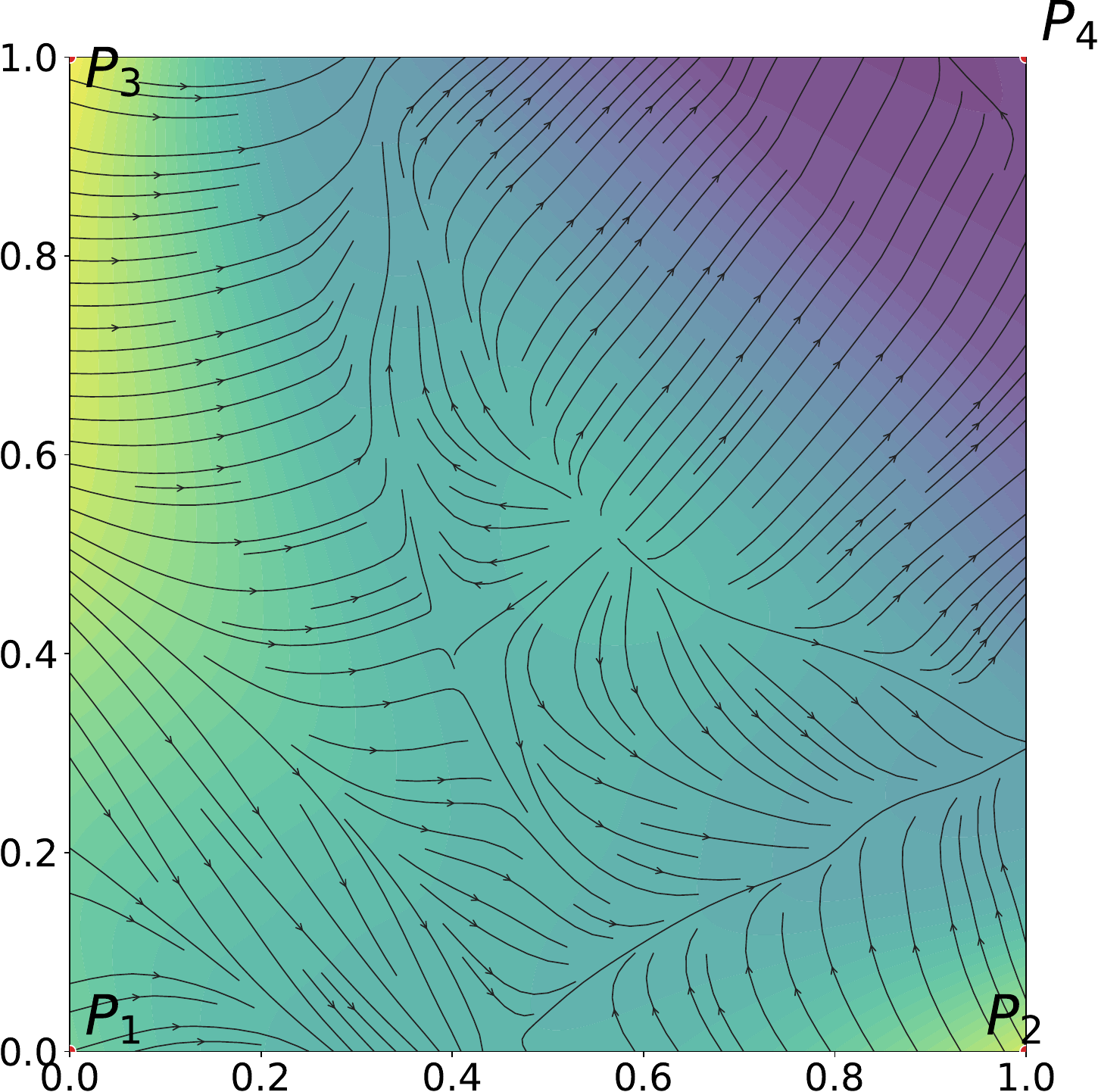} 
    \caption{14 tp} 
    \label{fig:cost_landscape:b}
  \end{subfigure}

  \caption{Contour plots of the cost landscape between four random plans for Wikidata star queries and the corresponding gradient flow. Even though the space between plans does not represent valid plans, the local gradient still contains directional information towards the better plan.}
  \label{fig:cost_landscape_all}
\end{figure}

\begin{figure}[h!]
  \centering
  \begin{subfigure}[b]{0.48\linewidth}
    \centering
    \includegraphics[width=\linewidth]{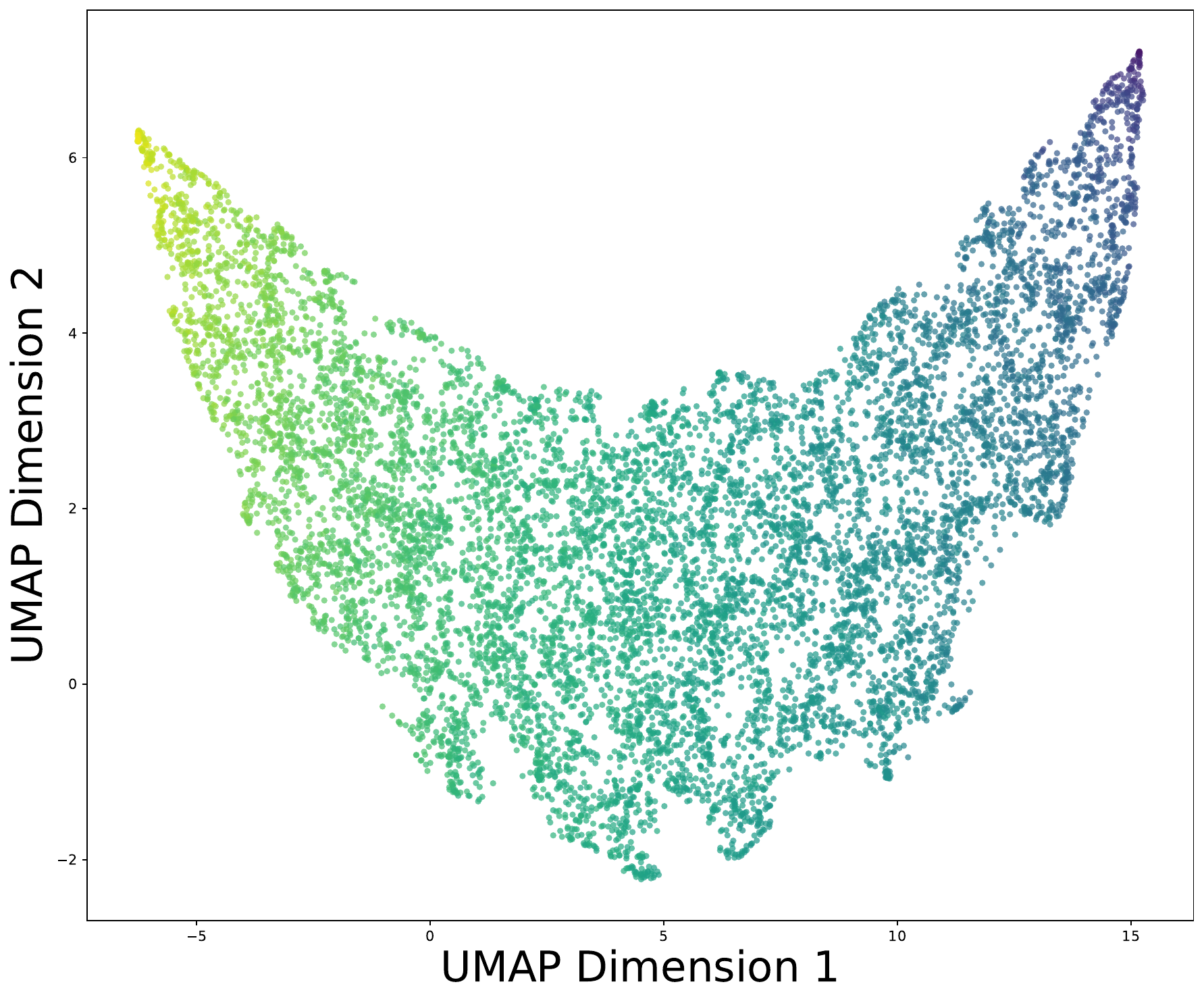}
    \caption{$\tau = 5$}   
    \label{fig:cost_landscape_umap:a}
  \end{subfigure}
  \hfill
  \begin{subfigure}[b]{0.48\linewidth}
    \centering
    \includegraphics[width=\linewidth]{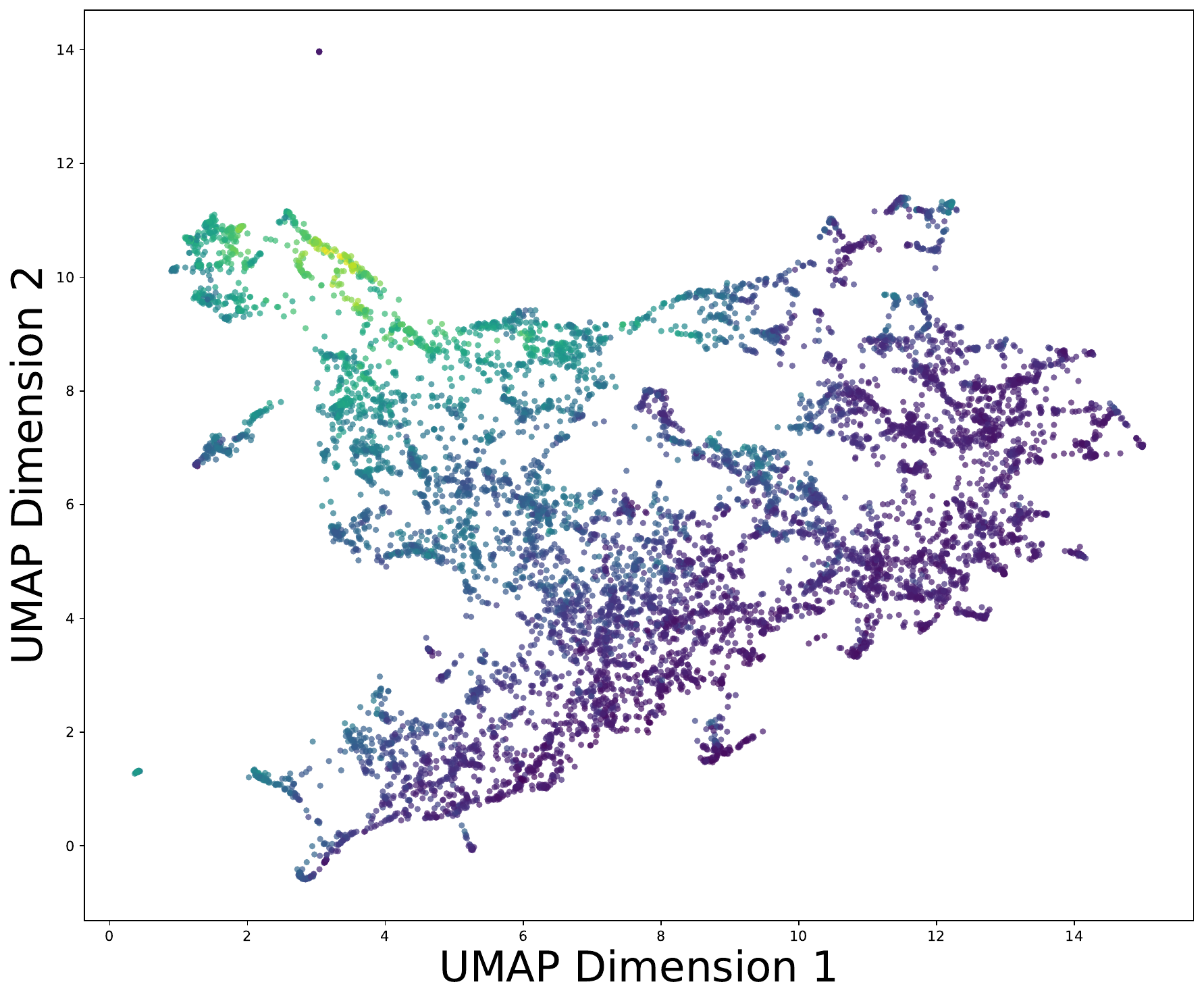} 
    \caption{$\tau = 0.01$} 
    \label{fig:cost_landscape_umap:b}
  \end{subfigure}

  \caption{UMAP Projection of soft plans for a single query and their predicted costs using the hidden representation of the Cost Model for different temperatures. For both high and low temperatures, costs smoothly change in the representation space.}
  \label{fig:cost_landscape_umap}
\end{figure}

To verify that the learned cost model produces a well-behaved landscape outside the set of discrete plans on which it was trained, we visualise the predicted cost landscape between four randomly chosen plans in Figure~\ref{fig:cost_landscape_all}. The plots show the predicted costs (yellow=high; purple=low) for a bilinear interpolation between the plans and the corresponding gradient flow. For the shown query with 8 triple patterns (Figure~\ref{fig:cost_landscape:a}), $P_4$ clearly has the lowest cost, and the relaxed landscape exhibits a rather monotonic gradient from all positions to a minimum near $P_4$. The query with 14 triple pattern in Figure~\ref{fig:cost_landscape:b} shows a more complex landscape. $P_3$ and $P_2$ have high cost, while $P_1$ now has medium cost. As a result, there exists a local maximum of cost in the center of the interpolation as well as (qualitatively) saddle points between $P_1, P_2$ and $P_2, P_4$. An optimizer with a small learning rate and without momentum may become trapped there. During the development of the approach, we indeed found that a higher learning rate and momentum significantly improve the quality of found plans. 
Since Figure~\ref{fig:cost_landscape_all} only shows a subspace of the full optimization space, we generated a more global view in Figure~\ref{fig:cost_landscape_umap}. For that, we sampled 10,000 different relaxed plans (with all incoming connections to triple patterns masked out) for a Wikidata star query with 5 triple patterns and represent each plan using the cost model's hidden representation of the plan (before the final graph-level MLP). We then generated 2D embeddings of those representations using UMAP. We show the same plans using a high (Figure~\ref{fig:cost_landscape_umap:a}) and a low (Figure~\ref{fig:cost_landscape_umap:b}) temperature. For $\tau=5$ (corresponding to the early phase of optimization), the representations are smoothly distributed and most importantly, monotonically distributed by cost. For $\tau=0.01$ (i.e., plans are approximately discrete), clusters that represent similar plans appear. However, there is still a clear trend from high to low cost. 

Overall, these results indicate that, even though the model has never been exposed to interpolated plans during training, the relaxed landscape provides meaningful directional information for gradient-based optimization.

\subsubsection{Comparison to Baseline Search Methods}
\begin{figure*}[t]
  \centering
  \begin{subfigure}[t]{0.245\textwidth}
    \centering
    \includegraphics[width=\linewidth]{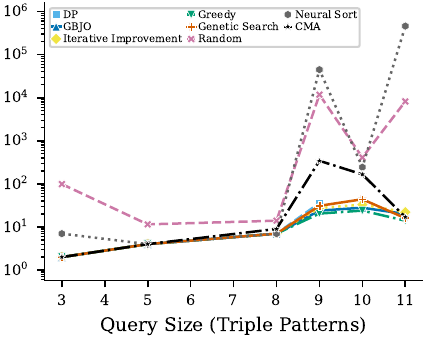}
    \caption{LUBM - Star Queries}
    \label{fig:optimization_results:a}
  \end{subfigure}\hfill
  \begin{subfigure}[t]{0.245\textwidth}
    \centering
    \includegraphics[width=\linewidth]{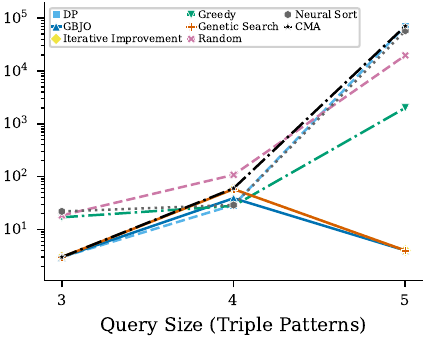}
    \caption{LUBM - Path Queries}
    \label{fig:optimization_results:b}
  \end{subfigure}\hfill
  \begin{subfigure}[t]{0.245\textwidth}
    \centering
    \includegraphics[width=\linewidth]{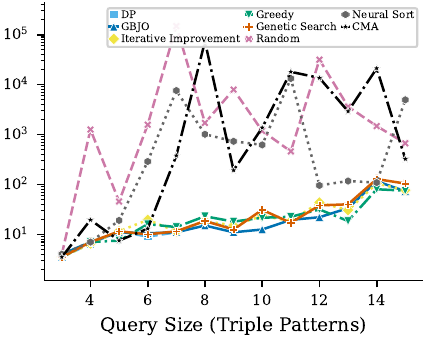}
    \caption{Wikidata - Star Queries}
    \label{fig:optimization_results:c}
  \end{subfigure}\hfill
  \begin{subfigure}[t]{0.245\textwidth}
    \centering
    \includegraphics[width=\linewidth]{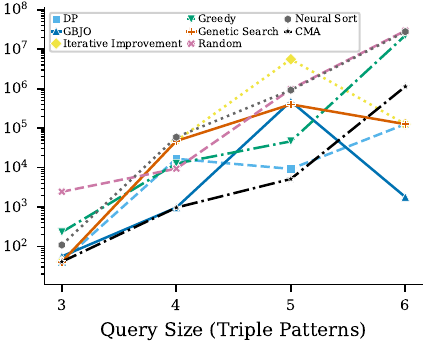}
    \caption{Wikidata - Path Queries}
    \label{fig:optimization_results:d}
  \end{subfigure}
  \caption{Median true costs ($y$-axis) of plans with respect to query size found by different join-order algorithms for LUBM and Wikidata.}
  \label{fig:optimization_results}
\end{figure*}

\begin{figure*}[t]
  \centering
  \begin{subfigure}[t]{0.245\textwidth}
    \centering
    \includegraphics[width=\linewidth]{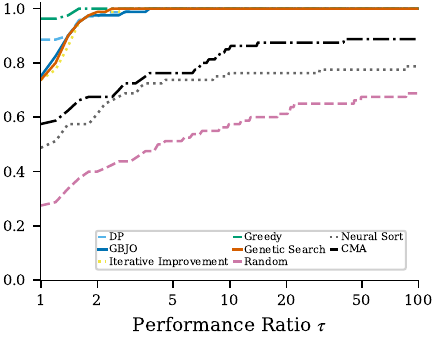}
    \caption{LUBM - Star Queries}
    \label{fig:dolan-more:a}
  \end{subfigure}\hfill
  \begin{subfigure}[t]{0.245\textwidth}
    \centering
    \includegraphics[width=\linewidth]{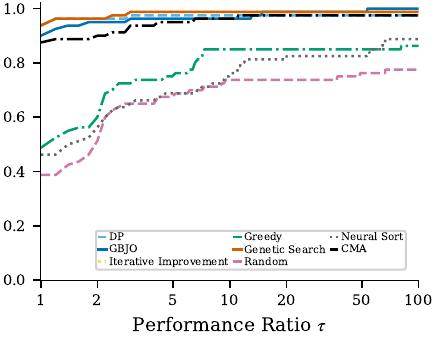}
    \caption{LUBM - Path Queries}
    \label{fig:dolan-more:b}
  \end{subfigure}\hfill
  \begin{subfigure}[t]{0.245\textwidth}
    \centering
    \includegraphics[width=\linewidth]{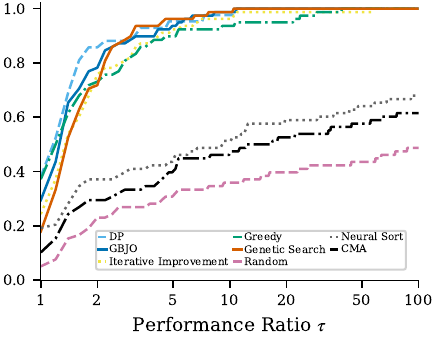}
    \caption{Wikidata - Star Queries}
    \label{fig:dolan-more:c}
  \end{subfigure}\hfill
  \begin{subfigure}[t]{0.245\textwidth}
    \centering
    \includegraphics[width=\linewidth]{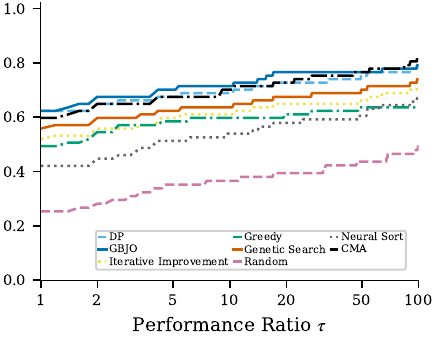}
    \caption{Wikidata - Path Queries}
    \label{fig:dolan-more:d}
  \end{subfigure}
  \caption{Dolan-Moré performance profiles of the compared join ordering algorithms showing the probability ($y$-axis) that a given method is within a fraction $\tau$ of the best solution found.}
  \label{fig:dolan_more}
\end{figure*}
Next, we evaluate how GBJO compares to other search methods. Figure~\ref{fig:optimization_results} compares the median true costs of plans across increasing query sizes and for both star- and path-shaped queries on LUBM and Wikidata. The results show that GBJO is competitive with the other discrete search methods, and sometimes even surpasses them. It is also robust with respect to growing query sizes. Compared to the costs of random plans, the costs of plans found by GBJO are, for large query sizes, several orders of magnitude lower. Again, this is a surprising result: although the relaxed space between discrete plans does not correspond to physically possible plans, it resembles interpolations between those (and their costs) and is informative to find (using the gradient of the cost) low-cost plans. Furthermore, in comparison to the other two approaches operating on the relaxed space, Neural Sort and CMA-ES, GBJO finds significantly lower costs. This shows that the specific combination of model prediction, temperature annealing, and time-dependent penalties (compared to Neural Sort), as well as using gradients to traverse the continuous plan space (compared to CMA-ES), is required to reach good results. Note again that since the cost model does not fulfill the Bellman optimality principle, DP does not necessarily yield the best found solutions (it is, however, still a strong baseline as the results show).

Figure~\ref{fig:dolan_more} displays the Dolan-Moré performance profiles for the different search methods on the four datasets. For each method, a Dolan-Moré profile shows the fraction of all queries for which this method is within a factor $\tau$ of the best-performing algorithm for each query. Hence, the y-intercept shows the percentage of cases where a given algorithm was the best among all others, while the tail to the right visualizes the robustness of an approach (the worst behavior compared to the best solution). In general, the higher and further to the left the curve of a given algorithm is, the better it performs. The performance plots show similar results compared to the median costs. GBJO has similar performance curves to the discrete baselines. Its robustness is also similar to that of the discrete baselines, showing that GBJO does not find plans significantly worse than the best found solution more often than the other approaches. 

Altogether, the results demonstrate that GBJO can match and even surpass commonly used discrete join ordering algorithms. GBJO can meaningfully traverse the continuous space of plan superpositions. This is intriguing since the cost model never saw plan superpositions during training, nor do they correspond to valid plans. Lastly, GBJO only uses $I=10$ evaluations of the cost model to guide the search as opposed to several hundred for the other randomized approaches (and Greedy Search for larger queries).
\subsection{Runtime Performance}
\begin{figure}[h!]
\centering
\includegraphics[width=0.49\textwidth]{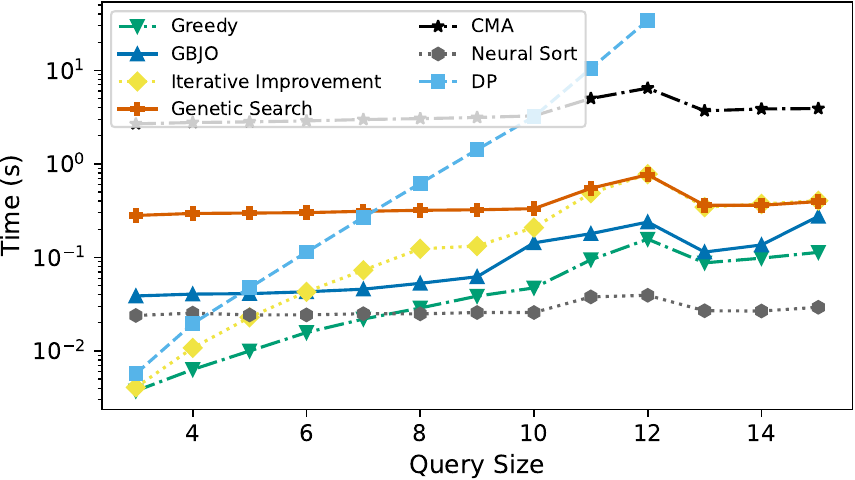}
\caption{Search runtimes of the compared baselines. GBJO has low runtime and is only dominated by Greedy Search and Neural Sort.}
\label{fig:runtime}
\end{figure}

Lastly, Figure~\ref{fig:runtime} shows the wall-clock time w.r.t. query size for GBJO and the compared algorithms on consumer-grade CPU.

As initially stated, DP exhibits an exponential increase in runtime. CMA-ES and GEQO have a higher and rather constant runtime due to the higher and constant number of steps. Iterative Improvement initially has a low runtime (as it quickly finds local minima), but its runtime quickly grows towards that of GEQO. Greedy Search, as expected, starts off with a very small runtime but has a steeper, quadratic trend. Due to the small number of required steps, GBJO also has a small runtime that is only dominated by Greedy Search and Neural Sort. Importantly, it exhibits a more favorable trend compared to Greedy Search. The absolute runtime between 40 and 200 ms makes GBJO a practical option for real-time join order optimization. Furthermore, since the current implementation is not runtime-optimized, substantial improvements in absolute time are expected by compiling the forward and backward passes into static graphs and employing additional runtime optimization techniques.

\section{Conclusion and Outlook}
We demonstrated the feasibility of approaching join order optimization using gradient descent by continuously relaxing the discrete search over query plan trees. Our results indicate that gradient-based search can successfully find similar or better plans compared to discrete search methods, and do so with favourable absolute runtime and complexity.

For future work, the runtime can be further enhanced using model compilation and efficient kernels for the presented penalties.

\bibliographystyle{named}
\bibliography{ijcai26}

\appendix
\section{Proof of Theorem 1}
\begin{theorem}
Let $n \geq 2$ and $A \in \{0,1\}^{N \times N}$ with $N=2n-1$, subject
to the masking introduced above: $A_{iv}=0$ for all $v \leq n$ and
$A_{rj}=0$ for all $j$, with $r = N$.
Then $P_{\textsc{to}}+P_{\textsc{ji}}+P_{\textsc{jo}}+P_{\textsc{ll}}
+P_{\textsc{acyc}} = 0$ if and only if $A$ is the adjacency matrix of
a valid left-linear plan.
\end{theorem}

\noindent\textit{Proof.}
($\Leftarrow$) A valid left-linear plan is a directed tree with all edges oriented
child$\to$parent, inducing a topological order that prevents closed walks;
hence $\operatorname{tr}(A^k){=}0$ for all $k{\geq}1$ and $P_\textsc{acyc}{=}0$.
The degree and left-linearity constraints hold by construction
($P_\textsc{to}{=}P_\textsc{ji}{=}P_\textsc{jo}{=}P_\textsc{ll}{=}0$).

($\Rightarrow$) Suppose $P_\textsc{struct}(A) = 0$.
\textit{(i)~Each penalty vanishes:}
$P_\textsc{to}, P_\textsc{ji}, P_\textsc{jo}, P_\textsc{ll}$ are sums of squares ($\geq 0$),
and $P_\textsc{acyc} = \operatorname{tr}(e^A) {-} N =
\sum_{k \geq 1} \operatorname{tr}(A^k)/k! \geq 0$, since each
$\operatorname{tr}(A^k) \geq 0$ for entry-wise non-negative $A$.
As $P_\textsc{struct}$ is a positively-weighted sum, each term is zero.
\textit{(ii)~Degree structure:}
Since $A$ is binary and $P_\textsc{acyc}{=}0$ excludes self-loops
($A_{vv}{=}1$ would give $\operatorname{tr}(A){\geq}1$),
$P_\textsc{to}{=}0$ gives $d_v^\text{out}{=}1$ for every
triple-pattern node $v {\leq} n$;
$P_\textsc{jo}{=}0$ gives $d_r^\text{out}{=}0$ and $d_v^\text{out}{=}1$ for non-root joins;
$P_\textsc{ji}{=}0$ gives $d_v^\text{in}{=}2$ for every join node;
masking gives $d_v^\text{in}{=}0$ for triple patterns.
Edge count: $|E| {=} n + (n{-}2) + 0 = N{-}1$.
\textit{(iii)~Acyclicity and connectivity:}
$P_\textsc{acyc}{=}0$ implies $\operatorname{tr}(A^k){=}0$ for all $k{\geq}1$.
Since $(A^k)_{ii}$ counts closed directed walks of length~$k$ from~$i$ to~$i$,
and any closed walk in a finite digraph contains a simple cycle,
$G(A)$ is acyclic.
Every non-root node has exactly one outgoing edge~(ii); following these edges yields a walk
that cannot revisit a node (acyclicity) and must terminate at the unique node with
out-degree~0, namely~$r$. Hence every node reaches~$r$, and $G(A)$ is a connected
graph on $N$ nodes with $N{-}1$ edges---i.e., a tree rooted at~$r$.
\textit{(iv)~Left-linearity:}
$P_\textsc{ll}{=}0$ gives $c_{j_0}^\text{tp}{=}2$, $c_{j_0}^\text{jn}{=}0$ for
the first join~$j_0$, and $c_v^\text{tp}{=}1$, $c_v^\text{jn}{=}1$ for every other
join~$v$. Since $j_0$ has no join child while every other join has exactly one,
the join nodes form a directed path $j_0 \to \cdots \to r$---exactly the
left-linear structure. \qed

\section{Cost Model and Training Details}
\subsection{Join and Triple Pattern Representation}
We represent a triple pattern $tp$ using an embedding vector $e_{tp}$. It consists of a concatenation of
the embeddings of the subject, predicate, and object of the triple pattern and an indicator dimension that this is a triple pattern node ($i=0$):
\[
e_{tp} = e_{s} \| e_{p} \| e_{o} \| i.
\]

For the individual embeddings, we follow the approach proposed by~\citeauthor{gnce} and use \emph{RDF2vec}~\cite{rdf2vec} embeddings. These work similarly to \emph{word2vec} embeddings and provide a high-dimensional semantic representation of the entity, encoding its neighborhood and function within the graph. We also add the \emph{count} $c$ of the entity to the embedding. The count represents the number of times the entity ($s$, $p$, or $o$) appears in the graph. This serves as a simple yet meaningful prior quantity to estimate cost, as it can represent an upper limit. In case an entity is a variable, all dimensions of the RDF2Vec embedding are set to 1. Finally, we prepend an indicator dimension $v$ to the full embedding to signal whether the entity is constant ($ v=0$) or variable ($v=1$). Altogether, we represent an entity $x$ as
\[
e_x = v \| e_{rdf2vec}(x) \| c.
\]
We choose 100-dimensional vectors for $rdf2vec$, hence a complete embedding for an entity is 102-dimensional. A full embedding for a triple pattern is then 307-dimensional. Join nodes carry no inherent information, but it is still important to provide initial embeddings so they can be sufficiently distinguished by the GNN. Those embeddings should purely be able to differentiate the join nodes, but do not encode any order (as the order is not predetermined for GBJO). Hence, for each query and every join node within, we sample a vector from a standard Gaussian distribution (normalized to unit length) matching the input dimension. This is then used as the initial representation of the join node.

\subsection{GNN Architecture}
The GNN $\hat{C}(X, A^{\mathrm{soft}})$ we use to transform a plan representation into a cost estimate consists of six message passing layers based on the GIN message passing function, which is extended by accepting edge weights to account for the soft adjacency matrix. The output $z_i^{(l)}$ for a node indexed with $i$ in layer $l$ is given by
{
\small
\[
\begin{aligned}
&\mathrm{GIN}(z_i^{(l-1)}) =  \mathrm{MLP}^{(l)}\!\Bigl((1+\epsilon^{(l)})\,h_i^{(l-1)} +\sum_{j\in\mathcal N(i)} a_{ji}^{\mathrm{soft}}\; h_j^{(l-1)} \Bigr)
\end{aligned}
\]
}
Here $\epsilon^{(l)}$ is a (learnable) scalar,  
$\mathcal N(i)$ the set of neighbours of node $i$, and  
$\mathrm{MLP}^{(l)}$ is a two–layer perceptron.
The message-passing layers are extended with skip connections and dropout (during training). A sum aggregation combines the final node embeddings, and 2 fully connected layers transform the aggregation into a cost estimate. The initial embeddings of triple patterns and join nodes are first transformed using a signed log1p function and then projected to the hidden dimension of the model. The signed log1p is important here as it brings the potentially large counts of the entities to a lower value. 
\[
\begin{aligned}
x^{(0)} &= W_{\mathrm{proj}} \cdot \mathrm{signlog1p}(x) \\
x^{(\ell)} &= x^{(\ell-1)} + \mathrm{GIN}^{(\ell)}(x^{(\ell-1)}, A^{\mathrm{soft}}, w) \\
x^{(\ell)} &= \mathrm{Dropout}(x^{(\ell)}) \\
h &= \sum_i x_i^{(L)} \\
g &= \mathrm{Dropout}(\mathrm{GELU}(W_1 h + b_1)) \\
\hat{C}(X,A^{\mathrm{soft}}) &= W_2 g + b_2
\end{aligned}
\]
\subsection{Training Procedure}
The GNN is implemented using Pytorch Geometric. One model is trained per dataset and query shape with as many training plans as mentioned in the experimental section. All models are trained for 200 epochs using a batch size of 128 with the AdamW optimizer using a learning rate of $10^{-4}$. As a loss function, the Huber loss between the predicted and true log-cost of a plan is used. The final model is chosen based on the lowest Q-Error on the validation plans.

\begin{algorithm}[t]
\caption{\textsc{Gradient-Based Join Ordering}}
\label{alg:gbjo}
\begin{algorithmic}[1]
\Require
  node features $X$, trained cost model $\hat C_\theta$,\\
  iterations $I$, learning rate $\alpha$, exponent $q$,\\
  temperatures $(\tau_0,\tau_{\min})$, penalty weights $\{\lambda_k\}$ 
\State Initialise logits $L_{ij} = 0$ \\
set $L_{ii}\leftarrow-\infty$
\State $(L^\star,C^\star)\gets(L,\infty)$
\For{$t=0$ \textbf{to} $I-1$}
    \State $\tau\leftarrow\max\bigl(\tau_{\min},\tau_0-\frac{t}{I}(\tau_0-\tau_{\min})\bigr)$
    \State $A^{\text{soft}}\leftarrow \mathrm{softmax} \bigl(L/\tau\bigr)$ 
    \State $\widehat c\leftarrow \hat C_\theta(X,A^{\text{soft}})$
    \State $P_{\textsc{struct}}\leftarrow\textsc{CalcPenalty}(A^{\text{soft}},\{\lambda_k\})$
    \State $\lambda(t)\leftarrow\lambda_{\max}\bigl(\tfrac{t}{I}\bigr)^q$
    \State $\mathcal L\leftarrow\widehat c+\lambda(t)P_{\textsc{struct}}$
    \State $L\leftarrow\text{Gradient Descent}(L,\nabla_L\mathcal L,\alpha, m)$
    \State $\widehat c \leftarrow \hat C_\theta(X,\textsc{ProjectDiscrete}(A^{\text{soft}}))$
    \If{$\widehat c<C^\star$}
        \State $(L^\star,C^\star)\gets(L,\widehat c)$
    \EndIf
\EndFor
\State \Return $p^\star\leftarrow \textsc{ProjectDiscrete}(A^{\text{soft}}(L^\star))$
\end{algorithmic}
\end{algorithm}

\begin{algorithm}[t]
\caption{\textsc{ProjectDiscrete} for Left-Linear Plans}
\label{alg:projectdiscrete_full}
\begin{algorithmic}[1]
\Require Soft adjacency $A^{\text{soft}}\in[0,1]^{N\times N}$, number of triple patterns $n$
\State $T_{\text{free}} \gets \{1,\dots,n\}$; $J_{\text{free}} \gets \{n+1,\dots,2n-1\}$
\State $c \gets 2n-1$
\While{$|T_{\text{free}}| > 2$}
  \State $(j^\star,k^\star) \gets \arg\max\limits_{j\in J_{\text{free}},\,k\in T_{\text{free}}}
      \left(A^{\text{soft}}_{jc} + A^{\text{soft}}_{kc}\right)$
  \State Add edges $(j^\star \to c)$ and $(k^\star \to c)$ to $p^\star$
  \State $c \gets j^\star$
  \State $T_{\text{free}} \gets T_{\text{free}} \setminus \{k^\star\}$
  \State $J_{\text{free}} \gets J_{\text{free}} \setminus \{j^\star\}$
\EndWhile
\State Let $k_1,k_2$ be the two remaining elements of $T_{\text{free}}$
\State Add edges $(k_1 \to c)$ and $(k_2 \to c)$ to $p^\star$
\State \Return $p^\star$
\end{algorithmic}
\end{algorithm}

\section{GBJO Details}
The full pseudocode of GBJO is given in Algorithm~\ref{alg:gbjo}. The pseudocode of how a soft adjacency matrix is transformed to a valid left-linear plan is presented in Algorithm~\ref{alg:projectdiscrete_full}.
\subsection{Hyperparameters}
The hyperparameters for GBJO on the different datasets are determined using a simple hyperparameter search over 400 plans using the default optimizer of Nevergrad~\cite{nevergrad}.
\begin{table}
  \centering
  \scriptsize
  \caption{Final hyperparameters used for gradient-based join ordering, determined using Bayesian hyperparameter search.}
  \label{tab:hpo_results}
  \begin{tabular}{lcccc}
    \toprule
    & LUBM-Star & LUBM-Path & Wikidata-Star & Wikidata-Path \\ \midrule
    $\alpha$            & 2.26   &  1.67  & 4.9 & 3.9 \\
    $\lambda_{\text{ACYC}}$          & 24.5   & 3.34 & 29 & 1.8 \\
    $\lambda_{\text{TO}}$      & 60.7 & 11.7 & 1.4 & 1.27 \\
    $\lambda_{\text{JI}}$         & 18.8   & 2.07 & 3.6 & 1 \\
    $\lambda_{\text{JO}}$        & 9.3   & 2.8 & 4.1 & 1.02 \\
    $\lambda_{\text{LL}}$        & 28.8   & 51.7 & 60 & 5.9 \\
    $q$     & 1.7   & 1.31   & 1.01  & 1.06 \\
    $\tau_0$    & 3.2   & 1.12   & 4  & 2.55 \\[2pt]
    $\tau_{min}$                 & 0.12   & 0.12  & 0.49  & 0.79 \\
    $\lambda_{\mathrm{max}}$     & 1   & 0.56   & 1  & 0.87 \\
    \bottomrule
  \end{tabular}
\end{table}
Table~\ref{tab:hpo_results} shows the final hyperparameters used for the different datasets. The absolute values of the different penalty coefficients vary significantly. However, most interestingly, a very high learning rate is important for all models. This is due to the fact that the optimization must converge near good plans with only 10 steps.

\end{document}